\documentclass[onecolumn,pra,superscriptaddress,showpacs,floatfix]{revtex4}
\usepackage{graphicx}
\usepackage{subfigure}
\usepackage[]{amsmath}
\usepackage{dcolumn}
\usepackage{bm}
\usepackage{tabularx}

\graphicspath{.}

\begin{document}

\title{Isotope shift in the Sulfur electron affinity: \\ observation and theory}

%
%
\author{Thomas Carette}
\thanks{PhD grant from the ``Fonds pour la formation \`a la Recherche dans l'Industrie et dans l'Agriculture'' of Belgium (Boursier F.R.S. - FNRS) }
\affiliation{Chimie Quantique et Photophysique, Universit\'e Libre de Bruxelles - CP160/09, B 1050 Brussels, Belgium}

\author{Cyril Drag}
\affiliation{Laboratoire Aim\'e-Cotton, CNRS, Universit\'e Paris-sud, F-91405 Orsay cedex, France}

\author{Oliver Scharf}
\affiliation{Chimie Quantique et Photophysique, Universit\'e Libre de Bruxelles - CP160/09, B 1050 Brussels, Belgium}

\author{Christophe Blondel} 
\affiliation{Laboratoire Aim\'e-Cotton, CNRS, Universit\'e Paris-sud, F-91405 Orsay cedex, France}

\author{Christian Delsart} 
\affiliation{Laboratoire Aim\'e-Cotton, CNRS, Universit\'e Paris-sud, F-91405 Orsay cedex, France}

\author{Charlotte {Froese Fischer}} 
\affiliation{National Institute of Standards and Technology 
             Gaithersburg, MD 20899-8420, USA}

\author{Michel Godefroid}
\email{mrgodef@ulb.ac.be}
\affiliation{Chimie Quantique et Photophysique, Universit\'e Libre de Bruxelles - CP160/09, B 1050 Brussels, Belgium} 

\date{\today}

\begin{abstract}
The electron affinities $^e\!A$(S) are measured for the two isotopes $^{32}$S and $^{34}$S (16752.9753(41) and 16752.9776(85)~cm$^{-1}$, respectively). The isotope shift in the electron affinity   is found to be positive, 
$^e\!A$($^{34}$S$)- \; ^e\!A$($^{32}$S) $ = + 0.0023(70)$~cm$^{-1}$, but the uncertainty allows for the possibility that it may be either ``normal''  [ $^e\!A$($^{34}$S) $ > \; ^e\!A$($^{32}$S)~]
 or ``anomalous'' [ $^e\!A$($^{34}$S) $ < \; ^e\!A$($^{32}$S)~]. The isotope shift is estimated theoretically using elaborate correlation models, monitoring the electron affinity and the mass polarization term expectation value. The theoretical analysis predicts a very large specific mass shift that counterbalances the normal mass shift and produces
 an anomalous isotope shift, $^e\!A$($^{34}$S$)- \; ^e\!A$($^{32}$S) $ = - 0.0053(24)$~cm$^{-1}$. 
The observed and theoretical residual isotope shifts agree with each other within the estimated uncertainties.
\end{abstract}

\pacs{31.30.Gs, 32.80.Gc, 31.15.ve, 32.10Hq }

\maketitle

\section{Introduction}

Photodetachment microscopy, which is the analysis of the electron interference pattern naturally produced when photodetachment occurs in the presence of an electric field \cite{Bloetal:96a,Bloetal:99a}, was applied to a beam of $^{32}$S$^-$ ions and allowed to measure the detachment thresholds corresponding to different fine-structure levels of the negative ion S$^-$ and the neutral atom S \cite{Bloetal:06a}. The electron affinity
of Sulfur, at 2.077~eV, is well suited for detachment by a tunable dye laser,
which provides a third way of measuring neutral~S fine structure, besides VUV
spectroscopy of S I lines and direct fine-structure resonance spectroscopy.
Dye laser photodetachment of S$^-$ was also used as a probe of
microwave-induced transitions of hyperfine-split Zeeman transitions,
which lead to a measurement of the hyperfine structure of $^{33}$S$^-$
\cite{Traetal:89a}.
The fine structures of S$^-$ and neutral~S, with the definition of the six ``fine-structure
detachment thresholds'', labelled A, B, C, D, E and F (in the order of increasing excitation energy)  are displayed in Figure~\ref{fig:FS_thresholds}. 
In the present work, photodetachment microscopy is used to measure the electron affinity $^e\!A$ (threshold C) of the even isotopes 32 and 34 of Sulfur. The sensitivity of the method made it possible to record a significant number of $^{34}$S detachment events, even though Sulfur was produced from a chemical compound with no isotopic enrichment and detachment occurred at very low energies above threshold (namely in the sub-meV range). The accuracy of photodetachment-microscopy based electron affinity measurements makes it possible to get an estimate of the isotope shift, precise enough to make comparison with theory significant. \\

On the theoretical side, the {\em ab initio} calculation of electron affinities is a challenge. Various methods for evaluating electron binding energies and affinities are discussed by Lindgren~\cite{Lin:05a}, presenting five different techniques, from the Koopmans-theorem method, up to the density-functional-theory. Lindgren's  survey focuses  on the many-body perturbation approach and only mentions, without further discussion, a sixth method that can always be used: separate many-body calculations for the
initial and final states, using some elaborate variational technique,
like multi-configuration  or configuration interaction methods. This approach can indeed be  successful for small systems (see for instance ref.~\cite{PacKom:06b} for $^7$Li) and is precisely the one attempted in the present work, although the correlation balance is more difficult to achieve for a large number of electrons. 

The electron affinity of Sulfur has been estimated, after the pioneer work of Clementi {\it et al.}~\cite{Cleetal:64a}, by Woon and Dunning~\cite{WooDun:93a} who treated the second row atoms
through multireference single and double excitation configuration interaction calculations and by Gutsev {\it et al.}~\cite{Gutetal:98a}  using the coupled-cluster method. 
In a benchmark study of {\it ab initio} and density-functional calculations of electron affinities covering the first- and second-row atoms, de Oliveira {\it et al.}~\cite{deOetal:99a} concluded that the best {\it ab initio} results agree, on average, to better than 0.001~eV with the most recent experimental results.
For heavy systems, the relativistic effects become crucial~\cite{Gar:95a,Guoetal:98a} and an accuracy better than 0.04~eV (10\%) was difficult to achieve for the electron affinity of lead in the full relativistic approach~\cite{Tatetal:09a}.

The search for a possible variation of the fine structure constant
$\alpha$ has renewed  interest in developing reliable {\it ab initio}
computational methods for atomic spectra~\cite{Beretal:08a,Poretal:09a}. 
Theory versus experiment comparisons of atomic isotope shifts 
can serve as sensitive tests of our computational ability for some important electronic factors.
Stimulated by photodetachment experiments~\cite{Beretal:95a,Valetal:99a}, theoretical calculations on isotope shifts in electron affinities have been attempted for Oxygen \cite{GodFro:99b,Bloetal:01a}, using the numerical MCHF approach. Beryllium was another interesting target~\cite{Nemetal:04a} but {\em ab initio} calculations remain scarce, requiring the calculation of two properties - the electron affinity {\it and} its isotope shift - both highly sensititive to correlation effects. For these systems, the limited population and the restricted active space concepts were used to build the configuration spaces. Adopting similar correlation models for the 
S/S$^-$ system becomes prohibitive and efficient reduction strategies need to be found.

In the present work, various single- and double-multireference valence expansions sets are explored for a full non-relativistic variational optimization of the wave function through the MCHF procedure.  Core excitations from these multireference sets are included afterward through configuration interaction. Computational strategies are  investigated and developed by monitoring the electron affinity and the mass polarization expectation values difference between the neutral atom and the negative ion. \\

The experimental work and theoretical calculations are presented in section~\ref{sec_exp} and \ref{sec_theory}, respectively. 
The comparison of observation and theory is discussed in section~\ref{sec_comp_obs_th}.

\section{Experimental measurement of the isotope shift} \label{sec_exp}
\subsection{Experimental setup}
\subsubsection{Ion beam and isotope selection}

Photodetachment microscopy was performed on a beam of S$^-$ produced by a hot cathode discharge in a mixture of 98\% Ar and 2\% CS$_2$. This commercial mixture had no isotopic enrichment, so the isotopes of Sulfur were produced with their natural abundance, which meant only 4\% \cite{NISTIAS} of $^{34}$S$^-$, {\it i.e.} only a few pA. With such a barely measurable ion current, all the electrostatic settings of the ion beam and the alignment of the laser in the interaction region had to be done with our Wien velocity filter set on mass 32. Then the electric field  applied in the Wien filter was reduced by the factor $4/\sqrt{17}$ appropriate for shifting from mass 32 to mass 34. We checked that the electron interferograms obtained with this setting were not a residue of $^{32}$S$^-$ signal on the wing of the maximum of mass-32 transmittance. A primary observation was that the peaks of the mass spectrum, when recorded on the total ion current signal (including a very visible contribution of SH$^-$ at mass 33) appeared well separated, which meant a mass resolution of 70 at least. A more quantitative limit of the possible admixture of the 32 signal at mass 34 was given by setting the velocity filter at mass 33, and observing that this actually let no photoelectron signal emerge from the background of the electron image, even though we recorded the impacts on the detector for a time longer than the one needed to reconstruct visible interference rings at mass 34. One mass unit away from its maximum, the $^{32}$S$^-$ current was thus much lower than the $^{34}$S$^-$ maximum current. Since isotope 32 is 22 times more abundant than isotope 34, this means a factor of attenuation of 100 at least. Two units away from the 32 maximum, the attenuation must be even more complete, so the few $^{32}$S$^-$ detachment events remaining at mass 34 are certainly negligible with respect to the electron background due to parasitic collisions of the ion beam with the diaphragms or the residual gas and to the 34 signal itself.

\subsubsection{Laser photodetachment}

As in previous photodetachment microscopy experiments on Sulfur \cite{Bloetal:05a,Bloetal:06a}, laser excitation was provided by a CW ring dye laser (Spectra-Physics 380A) operating with Rhodamine 590 in 5\% methanol and 95\% ethylene-glycol. Single-mode operation was achieved by means of a pair of intracavity Fabry-Perot etalons. The cavity length is servo-locked by means of an external sigmameter \cite{JunPin:75a}. Thanks to stabilization of the sigmameter itself on the wavelength of a dual-polarization stabilized He-Ne laser, the frequency of the tunable laser can remain stable within a few MHz for the typically 20 minutes needed to record every photoelectron interferogram. The wave-number of the laser is measured by an 
\AA ngstrom WS-U lambdameter, with an accuracy better than 10$^{-3}$~cm$^{-1}$.

\subsection{Experimental data}

\subsubsection{Photodetachment interferograms}

Figure \ref{fig:interferogram} gives an example of a pair of interferograms obtained from a double pass of the laser beam on the ion beam. This double pass makes it possible to obtain Doppler-free measurements, by averaging the responses of both spots, for they correspond to symmetric residual deviations from $90^\circ$ of the laser-ion intersection angle ({\it cf.}~\cite{Bloetal:01b}). Due to the rarity of $^{34}$S, even after an accumulation time of 2000~s, the number of electrons counted per pixel is 8 at its maximum. Nevertheless this is enough for the fitting program to find the centre and contour of each spot, and to calculate a histogram of the average number of electrons counted per pixel at a given distance from the centre. The obtained radial profile, as shown on Figure~\ref{fig:Radprof}, gives a much less noisy picture of the interference pattern and shows its excellent correspondence with theory (even though the actual fitting procedure is done on the 2D electron distribution). The phase, {\it i.e.} $2\pi$ times the number of oscillations in the interferogram, is the essential parameter for determining the initial kinetic energy of the electron. The interferometric accuracy so obtained, of the order of a few 10$^{-3}$~cm$^{-1}$, is orders of magnitude better than what a measurement of the spot diameter would provide. The latter would actually pertain to the domain of classical electron spectrometry, the accuracy of which is seldom better than 1~meV, or a few cm$^{-1}$.

\subsubsection{Data analysis}

In principle, photodetachment microscopy does not require series of photodetachment images to be recorded to get a measure of the electron affinity $^e\!A$. Calculating the difference between the photon energy and the measured electron kinetic energy in a single experiment would give the result. However, any discrepancy between the expected and actual values of the electric field in the photodetachment region is able to produce a systematic shift of the measured photoelectron energies. This being a constant relative error, extrapolating the measured electron affinity down to zero initial kinetic energy, {\it i.e.} to the detachment threshold, provides a method for avoiding the electric field uncertainty \cite{Goletal:05a,Bloetal:05a}. Series of measured electron affinities obtained both for isotope 32 and isotope 34 are represented on Figure~\ref{fig:donnees}.

Extrapolation of the measured $^e\!A$ values down to zero being the leading idea, we had to admit that the few experimental points obtained would not be enough to determine the slope of the linear regression, for the $^{34}$S case, with a satisfying accuracy. The idea was thus to make $^{32}$S and $^{34}$S measurements in similar experimental conditions, and make the linear regression with a constraint of similarity on the slopes, in order to set the $^{34}$S slope with improved accuracy. The experimental data shown on Figure~\ref{fig:donnees} for $^{32}$S were actually taken during the same runs as the $^{34}$S ones. The difference between the $^{34}$S and $^{32}$S slopes may thus be constrained by a normal distribution with a characteristic width of 0.5\%, which is an estimate (on the larger side) of the typical slope variations observed in past experiments done in similar conditions. As a matter of fact, fitting the data with this constraint yields nearly identical slopes of 0.18 \% and 0.16 \% for $^{32}$S and $^{34}$S respectively, the visual consequence of this result being that the two regression lines drawn on Figure~\ref{fig:donnees} appear nearly parallel.

\subsubsection{Experimental results}
\label{subsubsec_exp_res}

The electron affinities $^e\!A$(S), which are the ordinates at zero energy of the two lines drawn on Figure~\ref{fig:donnees}, are 16752.9753(41) and 16752.9776(85) cm$^{-1}$ for $^{32}$S and $^{34}$S respectively. The error bars given here take the statistical distribution of the data into account at a 2$\sigma$ level together with a possible $\pm 10^{-3}$ cm$^{-1}$ systematic error on wavenumber measurements.
The value of $^e\!A$($^{32}$S) incorporates all our previous work that
led to the published result of 16752.9760(42)~cm$^{-1}$ \cite{Bloetal:06a}.
The subsequent discovery that transverse magnetic field effects were
actually negligible \cite{Chaietal:08a} was applied to these former data, but
accounts only for $-$0.0001 of the $-$0.0007 revision of the most probable
value of $^e\!A$($^{32}$S) down to 16752.9753(41)~cm$^{-1}$; $^e\!A$($^{32}$S) remains the most accurately known of all electron affinities.

The isotope shift $^e\!A$($^{34}$S)-$^e\!A$($^{32}$S) is found to be $+0.0023(70)$ cm$^{-1}$. Having a better accuracy on the difference than on the least-well known of both electron affinities is a logical consequence of the experimental method. The direct comparison of the apparent electron affinities at similar detachment energies naturally provides a good accuracy on the isotope shift, even though the linear regression to the actual value of $^e\!A$ suffers from additional experimental unknowns. The strong covariance of the electron affinities is reinforced by the inclusion of the possible systematic error on wavelength measurements, which is, by definition, the same in both cases. Numerically, the final correlation found between the obtained variances of the electron affinities is $+0.57$.


\section{Theory and {\em ab initio} calculations}
 \label{sec_theory}

 \subsection{Theoretical Isotope Shift}

Adopting the ($A' > A$) convention where $A$ is the mass number, the isotope shift on the electron affinity defined as 
\begin{equation}
IS(A',A) = \delta  \; ^e \! A \equiv \; ^e \! A(A') - \; ^e \! A (A) 
\end{equation}
 is expressed as the sum of the normal mass shift (NMS), specific mass shift (SMS) and field shift (FS) contributions
\begin{equation}
IS(A',A) = 
\delta ^e \! A_{\text{\sc{nms}}} 
+ \delta ^e \! A_{\text{\sc{sms}}}
+ \delta ^e \! A_{\text{\sc{fs}}} \; .
\end{equation}
Introducing $M$ for the nuclear mass and $X$ for the chemical element, the two first terms that constitute the mass shift  can be written in atomic units $(m_e = 1 $ and $ \hbar = 1)$ as
\begin{equation}
\label{NMS_plus_SMS}
\delta ^e \! A_{\text{\sc{nms}}} 
+ \delta ^e \! A_{\text{\sc{sms}}} = 
\left[ \frac{M'}{1+M'} - \frac{M}{1+M}\right]\ ^e\!A(\infty ) + \left[ \frac{M'}{(1+M')^2} - \frac{M}{(1+M)^2}\right]\ \Delta S_{\text{\sc{sms}}}
\end{equation}
 where
\begin{equation}
\Delta S_{\text{\sc{sms}}} =   S_{\text{\sc{sms}}} (X) -  S_{\text{\sc{sms}}} (X^-) \; ,
\end{equation}
with
\begin{equation}
S_{\text{\sc{sms}}} = 
- \left\langle \Psi_\infty  \left| \sum_{i<j}^N \nabla_i \cdot \nabla_j \right| \Psi_\infty  \right\rangle \; . 
\end{equation}
This expression is correct to first order in $\mu / M$, where $\mu = m_e M/(m_e + M)$ is the reduced mass of the electron with respect to the nucleus.  For a positive $\Delta S_{\text{\sc{sms}}}$ difference, the NMS and SMS interfer negatively due to the relative signs of the mass factors in equation~(\ref{NMS_plus_SMS})~\cite{Nemetal:04a}. It is easy to show that 
the degree of cancellation between NMS and SMS is basically governed by the mass-independent difference 
$ 
[ ^e\!A(\infty ) - \frac{\hbar^2}{m_e}\Delta S_{\text{\sc{sms}}}] \; . 
$
The two atomic masses of $^{32}$S (31.972~071~00~u) and $^{34}$S (33.967~866~90~u), taken from the AME2003 compilation of Audi {\it et al.}~\cite{Audetal:93a} are converted into nuclear masses by subtracting the electron mass contribution (0.027 \%)\footnote{The mass-equivalent of the electron binding energy is three orders of magnitude smaller than the electron mass contribution.}.

The field shift (FS) can be estimated from
\begin{equation}
\label{eq:FS}
\delta ^e \! A_{\text{\sc{fs}}} = (hc)
4\pi \left[ \rho({\bf 0})_{NR}^{X} - \rho({\bf 0})_{NR}^{X^-} \right]
\frac{ a_0^3}{4Z} f(Z)^{AA'}
 \left[  \langle r^2 \rangle_{A'} - \langle r^2 \rangle_{A}   \right]
\end{equation}
where $\rho({\bf 0})_{NR}$ is the non-relativistic spin-less total electron density $\rho({\bf r})$ (in $a_0^{-3}$) calculated at 
\mbox{${\bf r} = {\bf 0}$}~\cite{Boretal:10a}. 
The factor $f(Z)^{32-34}= 0.014823~\mbox{cm}^{-1}$/fm$^2$ is taken from Aufmuth {\it et al.}~\cite{Aufetal:87a} and corrects for the fact that we use the 
non-relativistic density for a point nucleus.
The $ \langle r^2 \rangle ^{1/2}_{32} = 3.2608(18)$~fm 
and $ \langle r^2 \rangle ^{1/2}_{34} = 3.2845(21)$~fm values are taken from Angeli~\cite{Ang:04a}.

Note from eq.~(\ref{eq:FS}) that with a positive variation of the rms nuclear charge radii, {\it i.e.} 
$\delta \langle r^2 \rangle ^{AA'} \equiv \langle r^2 \rangle_{A'} - \langle r^2 \rangle_{A} \geq 0 $, the FS has the same sign as the NMS if and only if the electron detachment ($X^- \rightarrow X $) is accompagnied by an increase of the electron density at the nucleus $ ( \Delta \rho({\bf 0}) \geq 0 ) $. For a system like Sulfur, the FS is expected to be much smaller than the mass shift. Therefore, our computational strategy is dictated by the description of the electron affinity and $\Delta S_{\text{\sc{sms}}}$, although the FS is taken into account in the present analysis (see section~\ref{sec_comp_obs_th}).

 \subsection{Computational method}

We use the numerical multi-configuration Hartree-Fock approach (MCHF) describing the atomic wave function as
\begin{equation}
\Psi = \sum_i c_i \Phi (\gamma_i LS)
\end{equation}
where  $ \{ \Phi (\gamma_i LS) \} $ is an orthonormal set of configuration state functions (CSF) that are symmetry adapted linear combinations of Slater determinants~\cite{Fro:77a,FBJ:97a}. In this method, the radial functions $\{ P_{nl}(r) \}$ defining the orbital active set and the mixing coefficients $\{c_i \}$ are variational. The configuration interaction (CI) method solves the eigenvalue problem in a CSF basis built with a fixed preoptimized orbital set.

For any differential property that is estimated from the difference between two calculated diagonal properties using the variational approach, much care must be taken to obtain a good balance between the two states. It becomes even more difficult when the latter belong to systems with different numbers of electrons. In such situations, the 
\textsc{atsp2k} package~\cite{Froetal:07a} is an efficient tool thanks to its flexibility. In particular, the fully implemented \emph{limited population} (LP)  and \emph{multi-reference} (MR) \cite{Godetal:96a,Jonetal:96a} approaches for building  the configuration space offer systematic ways of including and monitoring correlation.
While the LP configuration space is built by allowing single (S)-, double (D)-, triple (T)- and possibly higher excitations, from a single reference configuration state function, restricted by orbital occupation, the MR method is usually limited to SD excitations from a larger set of configuration states. 
 The LP method has been successfully used to calculate the isotope shifts in the Oxygen electron affinity~\cite{GodFro:99b,Bloetal:01a}. Although both the LP and MR correlation models have been explored, the present theoretical discussion is limited to the multireference calculations.

 \subsection{The experimental electron affinity as a guideline} \label{sub_EA_guideline}
 
  In our approach, the experimental electron affinity is used as a guideline to set efficient pathways in the variational configuration spaces. Our non-relativistic approach targets electron correlation. In this context, it is useful to get a reference non-relativistic $^e\!A$ value. 
Introducing the observed average energy levels
\begin{equation}
\label{averaging_J}
\overline{E}  = \frac{\sum_{J} (2J+1) E_J}{\sum_{J} (2J+1)} \; ,
\end{equation}
for both S $3p^4 \; ^3P$ and S$^-$~$3p^5 \; ^2P^o$,
the average experimental electron affinity that would be measured if not resolving the fine structure thresholds, is estimated from 
\begin{equation}
^e\!A_{exp}^{AV} = \overline{E}(\mbox{S} \; ^3P) - \overline{E} (\mbox{S}^- \; ^2P^\circ) =
\frac{(E_0 + 3 E_1 + 5 E_2)}{9} 
                 - \frac{(2 E_{1/2} + 4 E_{3/2})}{6} \; .
\end{equation}
This average energy can be rewritten 
in terms of some deviation to the observed electron affinity $ ^e\!A_{exp} = (E_2 - E_{3/2})$ (arrow $C$ in Figure~\ref{fig:FS_thresholds}) 
\begin{equation}
\label{eq:E_exp_av}
^e\!A_{exp}^{AV} =  \; ^e\!A_{exp}  + \frac{ 3(E_1 - E_2) + (E_0 - E_2)}{9} 
                                    - \frac{( E_{1/2} - E_{3/2})}{3}  \; .
\end{equation}
Using $^e\!A_{exp} = 1675 2.9753(41)$~cm$^{-1}$ reported in section~\ref{subsubsec_exp_res} for $^{32}$S and the  observed fine structures of S and S$^-$~\cite{Bloetal:06a}, the Sulfur average experimental electron affinity 
$
 ^e\!A_{exp}^{AV} = 16787.55~\mbox{cm}^{-1} = 2.081~391~\mbox{eV} = 0.076~489~702~\mbox{E}_{\mbox{h}} 
$
is obtained from eq.~(\ref{eq:E_exp_av}).

When adopting a Breit-Pauli description~\cite{Hibetal:91a,FBJ:97a} of atomic structures, the total binding energy $E(LSJ)$ of a level is expressed in first order perturbation theory as
\begin{equation}
\label{NF_F_partition}
E(LSJ) = E^{NR}_{LS} +  E^{NF}_{LS} + E^{F}_{LSJ} \; ,
\end{equation}
{\it i.e.} as the summation of the non-relativistic total energy $(E^{NR})$, the non-fine structure relativistic shift $(E^{NF})$ and the $J$-dependent fine-structure correction $(E^{F})$. 
Since the fine structures of both S $3p^4 \; ^3P$ and S$^-$~$3p^5 \; ^2P^o$ are washed out by the averaging process~(\ref{eq:E_exp_av}) to get $^e\!A_{exp}^{AV}$, a reference non-relativistic electron-affinity $^e\!A_{ref}^{NR}$ is estimated by subtracting the corresponding theoretical non-fine structure contribution 
 $\Delta E^{NF} = E^{NF}(\mbox{S}) - E^{NF}(\mbox{S}^-)$, from  the above experimental average electron affinity
\begin{equation}
\label{eq:E_exp_nr}
 ^e\!A_{ref}^{NR} =  \; ^e\!A_{exp}^{AV}  - \Delta E^{NF} \; .
\end{equation} 
The non-fine structure contribution calculated in the single configuration Hartree-Fock approximation, 
$\Delta E_{HF}^{NF} =  - 5.36251~10^{-4}~\mbox{E}_{\mbox{h}}$,
produces  a  non-relativistic electron-affinity value of $ ^e\!A_{ref}^{NR} = 0.077~026~\mbox{E}_{\mbox{h}} $.
Note that the latter value is in line with the estimation of the ``non-relativistic experimental'' electron-affinity ($0.076~939~\mbox{E}_{\mbox{h}}$) calculated from the scalar contribution 
of de Oliveira {\it et al.}~\cite{deOetal:99a}, who found an excellent general agreement for electron affinities of first- and second-row atoms.

\subsection{The Multi-Reference approach}

\subsubsection{MR-MCHF calculations}
\label{subsub_MR-MCHF}

In the multi-reference approach, one first defines a zeroth-order set of CSFs labelled MR

\begin{equation} 
\text{MR} \equiv \{ \Phi_1 (\gamma_1 LS\pi), \Phi_2 (\gamma_2 LS\pi), \ldots , \Phi_m (\gamma_m LS\pi) \} 
\end{equation}
that includes the dominant interacting terms in the description of a given atomic state. 
This zeroth-order set is then expanded to capture major correlation effects.  Useful expansions are built by allowing all single and double excitations from a multi-reference (MR-SD) set within a given orbital active space. 
From a practical point of view, these expansions are generated using \textsc{lsgen}~\cite{StuFro:93a} that produces the desired list of configurations, containing the complete set of CSFs for a given $LS\pi$~- total symmetry.

A good valence correlation MR-SD expansion for S$^-$ is based on the multireference set
 \begin{equation}
 \label{mr_s_moins}
\text{MR(S}^-)=\{1,2\}^{10} \{3s, 3p\}^5 \{3,4\}^2 \; .
\end{equation}
The notation is inspired from the LP approach: the multireference set (\ref{mr_s_moins}) is composed of all CSFs having the required symmetry (here $\; ^2P^o$), 
with ten electrons $ \{ 1,2 \}^{10} $ forced to occupy the $n=1$ and $n=2$ shells ({\it i.e.} a $ 1s^2 2s^2 2p^6 $ closed core). 
In the MR space, the seven valence electrons should describe the dominant configuration $3s^2 3p^5$, but they are also free to reorganize themselves in the $n=3$ and $n=4$ subshells with only one occupation constraint: a minimum of five electrons should be either $3s$, or $3p$, as explicitly stated through the notation $\{3s, 3p \}^5$.
But even with a closed core $\{1,2\}^{10}$, the computational effort is gigantic. 
Introducing the $\lceil n_{max} l_{max} \rceil$ notation for the orbital active set, 
the size of the expansion generated with six correlation layers (MR-SD$\lceil 9k \rceil$) reaches
1~895~416~CSFs.
Moreover, such a strategy is not efficient, a large number of  components being negligible. 
An interesting approach uses the ``Multi-Reference Interacting''(MR-I) CSF-space  defined as the union of the original set of CSFs that belong to the MR, and all CSFs that directly interact with at least one component of the MR, {\it i.e.} 
\begin{equation}
\label{MR-I}
\Phi_i(\gamma_i\ LS) \in \text{MR-I} \Leftrightarrow \exists \Phi_k \in \text{MR with } \left\langle \Phi_i(\gamma_i\ LS) \left| H \right| \Phi_k(\gamma_k\ LS) \right\rangle \neq 0 \; ,
\end{equation}
for any one-electron radial function basis set $\{P_{nl}(r) \}$.
 This selection constraint depends on the coupling ordering of the subshells. The conventional coupling hierarchy is a sequential one corresponding to the coupling of each subshell angular momenta to the previous intermediate coupling angular momenta, from left to right~\cite{Fan:65a}, for the natural subshell ordering ($n$ and $l$ increasing). But this is not always the most efficient representation. It is indeed common knowledge that the most strongly interacting momenta should be coupled first to get the best physical picture of the resulting levels pattern.  On this basis, we use the {\it reverse} order of orbitals, coupling sequentially the subshells by decreasing $n$ and $l$.  
 
Like the MR-SD space, the \mbox{MR-I} configuration set includes at most double excitations with respect to the reference but the building rule~(\ref{MR-I})  reduces drastically the size of the expansions in comparison with the MR-SD sets.
For example, the list of 
1~895~416~CSFs  discussed above is reduced to 525~111~CSFs in the MR-I$\lceil 9k \rceil$ model. The corresponding {\it configuration} reduction is much smaller ($ 18~576 \rightarrow 13~973$) since this reduction only arises from the intermediate coupling constraints associated with (\ref{MR-I}) and not from additional orbital occupation number selection rules. 
From a practical point of view, the MR-SD expansions are reduced according to the building rule~(\ref{MR-I}) to produce the desired MR-I lists using the 
\textsc{lsreduce} code integrated in the  \textsc{atsp2k} package~\cite{Froetal:07a}.

Similar to (\ref{mr_s_moins}), we explore the MR set
 \begin{equation}
\label{mr_s_1}
\text{MR1(S}) = \{1,2\}^{10} \{3s, 3p \}^4 \{3\}^2 \; ,
\end{equation}
for S. To discuss the delicate balance between the negative ion and the neutral atom, we  introduce for the latter a second  model, more correlated, based on the following MR set
\begin{equation}
\label{mr_s_2}
\text{MR2(S})= \{1,2\}^{10} \{3s, 3p \}^4 \{3\}^1\{3,4\}^1 \; .
\end{equation}
The orbital active sets are extended up to $n=9$, limited to $l\leq 7$ ({\it i.e.} $k$-orbitals).    
We use multireference sets with at most two excitations in the $3d$ subshell.
Allowing more than one $3d$ electron is indeed necessary for all important intermediate couplings to appear in the set of CSFs satisfying~(\ref{MR-I}).

  The MR set (\ref{mr_s_moins}) contains 157 CSFs. Some sublists of this complete configuration space are selected according to their impact on the energy, mass polarization and density at the nucleus and used to check the consistency of our results. By investigating the impact of a seventh correlation layer (10k) using  multireference subsets for S$^-$, this extra layer is estimated to contribute less than 5.$10^{-5}$ E$_h$ to both $^e\!A$ and $\frac{\hbar^2}{m_e} \Delta$S$_{\text{\sc{sms}}}$. 

With the valence correlation expansions based on MRs~(\ref{mr_s_moins}), (\ref{mr_s_1}) and (\ref{mr_s_2}), we choose to vary all orbitals in the MCHF approach, the frozen-core approximation being considered {\it a priori} artificial. 
The mean radii of the spectroscopic orbitals of S and S$^-$ are reported in Table~\ref{rOrb} and compared with the HF ones. 
As expected, we observe an overall stability of the $n=2$ orbitals and  a larger variation for the ($n=3$) valence shells, while the $1s$ orbital remains very similar in all calculations.
\begin{table}
\caption{Mean radius of spectroscopic orbitals in atomic units of length (a$_0$) obtained for S$^-$ and S with MR1 and MR2 including 6 correlation layers $(9k)$.\label{rOrb}}
\begin{tabular}{lcccc|ccc}
\colrule
\multicolumn{1}{c|}{}  & \multicolumn{4}{c|}{S $3p^4 \; ^3P$} & \multicolumn{3}{c}{S$^-$ $3p^5 \; ^2P^o$}\\
\multicolumn{1}{c|}{$nl$\hspace*{0.5cm}} &&\multicolumn{1}{c}{\hspace*{0.5cm} HF                    \hspace*{0.5cm}}& 
       \multicolumn{1}{c}{\hspace*{0.5cm}MR1-I$\lceil 9k \rceil$ \hspace*{0.5cm}} & 
       \multicolumn{1}{c|}{\hspace*{0.5cm}MR2-I$\lceil 9k \rceil$ \hspace*{0.5cm}}&& 
       \multicolumn{1}{c}{\hspace*{0.5cm} HF \hspace*{0.5cm}} & 
       MR-I$\lceil 9k \rceil$\\
\colrule
\multicolumn{1}{c|}{$1s$\hspace*{0.5cm}} && $0.09715$ & 0.09715 & 0.09715  && 0.09715 & 0.09715\\
\multicolumn{1}{c|}{$2s$\hspace*{0.5cm}} && 0.47577 & 0.47245 & 0.47248  && 0.47585 & 0.47276\\
\multicolumn{1}{c|}{$2p$\hspace*{0.5cm}} && 	0.44104 & 0.44061 & 0.44061  && 0.44106 & 0.44037\\
\multicolumn{1}{c|}{$3s$\hspace*{0.5cm}}&& 1.72072 & 1.71439 & 1.61071  && 1.77672 & 1.77863\\
\multicolumn{1}{c|}{$3p$\hspace*{0.5cm}}&& 2.06072 & 2.02942 & 1.84593  && 2.32369 & 2.25355\\
\colrule
\end{tabular}
\end{table}
Even if the $( 1s^2 2s^2 2p^6 )$ core is kept closed in the MCHF expansions, some correlation orbitals extend into the inner region of the atom to improve the description of the total wave function. In the Sulfur MR2-I$\lceil 9k \rceil$ calculation for example, the MCHF optimization involves 39 numerical correlation orbitals
from which the resulting $6p, 7d, 8s$ and $9f $ functions can be qualified as ``inner'' orbitals by looking at their mean radius \mbox{($ \langle r\rangle_{nl} < $ a$_0$)}. Although they still describe the (inner region) valence correlation, they lie in the correct region for estimating core-valence correlation effects through configuration interaction, as presented in the next subsection.


\subsubsection{Open-core configuration interaction calculations}

We add to the valence configuration lists, core-valence  mono- and multi-reference SD expansions (MR-CV-SD)  created by allowing at most one hole in the core but keeping the $1s$ shell closed and inactive. 
Core-valence excitations generate much larger lists of configurations than equivalent valence expansions. 

For keeping the size of the expansions tractable we use the following procedure. 
First we sort the configurations of the original MR-I valence eigenvectors into decreasing order
by their configuration weights. The latter is defined as the weighted contribution of the CSFs belonging to it:
\begin{equation}
w = \sqrt{\sum_{\Phi_i \in \{config\}} c_i^2} \; ,
\end{equation}
and are reported in Table~\ref{tab:weights}, for S$^-$ and for the two valence models (MR1 and MR2) used for S.
Following this hierarchy,  we define $p$ reference subsets MR$_p$  containing the first $p$ configurations in the sorted list.
Secondly, we build the corresponding MR$_p$-CV-SD spaces and keep only the CSFs interacting directly with the complete MR. We denote unambiguously the open-core CI calculations - all performed with six correlation layers - MR-I/CV$\! p$.

\begin{table}[h!]
\caption{Weights ($w$) of the configurations composing MR1, MR2 of S and the MR of S$^-$ in the corresponding MR-I$\lceil 9k\rceil$ wave-functions. \# denotes the configuration index.\label{tab:weights}}
\begin{tabular}{lclc|r|lcc|r|lc}
\colrule
 \multicolumn{4}{c|}{S $3p^4 \; ^3P$} & \multicolumn{7}{c}{S$^-$ $3p^5 \; ^2P^o$}\\
 \colrule
 \multicolumn{2}{c}{MR1-I$\lceil 9k\rceil $}  &  \multicolumn{2}{c|}{MR2-I$\lceil 9k\rceil $} &  \multicolumn{7}{c}{MR-I$\lceil 9k\rceil $}\\
config. & $w$ & config. & $w$ & \# & config. & $w$ && \# & config. & $w$\\
 \colrule
$3s^2 3p^4 $&$ 0.9567 $&$ 3s^2 3p^4 $&$ 0.9151 $&$ 1 $&$ 3s^2 3p^5 $&$ 0.9382 $&$\; $&$ 16$&$ 3p^5 3d^2 $& 0.0296\\
$3s 3p^4 3d $&$ 0.1605 $&$ 3s^2 3p^3 4p $&$ 0.2652 $&$ 2 $&$ 3s^2 3p^3 3d^2 $&$ 0.1773 $&$ $&$17$&$ 3s^2 3p^3 3d 4s $& 0.0292\\
$3s^2 3p^2 3d^2 $&$ 0.1568 $&$ 3s 3p^4 3d $&$ 0.1573 $&$ 3 $&$ 3s^2 3p^3 4p^2 $&$ 0.1249 $&$ $&$18$&$ 3s^2 3p^3 4s^2 $& 0.0291\\
$3p^4 3d^2        $&$ 0.0498 $&$ 3s^2 3p^2 3d^2 $& 0.1508 
&$ 4$&$ 3s 3p^5 3d $&$ 0.1045  $&$ $&$19$&$ 3p^5 3d 4d $& 0.0288\\
&$ $&$ 3s 3p^4 4s $& 0.1256
&$5$&$ 3s^2 3p^4 4p $&$ 0.1030  $&$ $&$ 20$&$ 3s^2 3p^3 4s 4d $& 0.0287\\
&$ $&$ 3s 3p^3 3d 4f $& 0.0655
&$6$&$ 3s 3p^4 4s 4p $&$ 0.0748  $&$ $&$ 21$&$ 3p^6 4p $& 0.0252\\
&$ $&$ 3s 3p^3 3d 4p $& 0.0632
&$7$&$ 3s^2 3p^3 3d 4d $&$ 0.0664 $&$ $&$ 22$&$ 3p^5 4p^2 $& 0.0239\\
&$ $&$ 3s^2 3p^3 4f $& 0.0531
&$8$&$ 3s 3p^4 3d 4f $&$ 0.0647 $&$ $&$ 23$&$ 3p^5 4d^2 $& 0.0203\\
&$ $&$ 3p^4 3d^2 $& 0.0492
&$9$&$ 3s 3p^4 3d 4p $&$ 0.0626 $&$ $&$ 24$&$ 3p^5 4s^2 $& 0.0200\\
&$ $&$  3p^5 4p $& 0.0310
&$10$&$ 3s 3p^4 4p 4d $&$ 0.0599 $&$ $&$ 25$&$ 3s^2 3p^3 4p 4f $& 0.0173\\
&$ $&$ 3s^2 3p^2 3d 4d $& 0.0247
&$11$&$ 3s^2 3p^4 4f $&$ 0.0537  $&$ $&$ 26$&$ 3p^5 4f^2 $& 0.0124\\
&$ $&$ 3p^4 3d 4s $& 0.0142
&$12$&$ 3s 3p^5 4d $&$ 0.0447  $&$ $&$ 27$&$ 3s 3p^5 4s $& 0.0074 \\
&$ $&$ 3s 3p^4 4d $& 0.0127
&$13$&$ 3s^2 3p^3 4d^2 $&$ 0.0439 $&$ $&$ 28$&$ 3p^5 3d 4s $& 0.0021\\
&$ $&$ 3p^4 3d 4d $& 0.0085
&$14$&$ 3s^2 3p^3 4f^2 $&$ 0.0401 $&$ $&$ 29$&$ 3p^5 4s 4d $& 0.0020\\
&$ $&$ 3s^2 3p^2 3d 4s $& 0.0030
&$15$&$ 3s 3p^4 4d 4f $&$ 0.0322  $&$ $&$ 30$&$ 3p^5 4p 4f $& 0.0018\\
&$$&$$&$$&$$&$ $&$ $&$ $&$ 31$&$ 3s 3p^4 4s 4f $& 0.0010\\
\colrule
\end{tabular}
\end{table}

Table \ref{tab:EA} reports the electron affinity theoretical values calculated with the MR (S$^-$) and  MR2 (S) models. 
 The largest   configuration interaction calculation  remains feasible  for Sulfur (MR2-I/CV15),
but the computational limits are definitely exceeded in the negative ion (MR-I/CV31). 
For S$^-$  indeed, a larger calculation than MR-I/CV14 requires truncating the expansions~\footnote{To perform this CI calculation (3~175~092 CSFs), we used 24 processors of 2.4 GHz during about 24 hours. The matrix  and whole set of useful integrals were stored on disk and took a space of 500 Gb. The memory necessary to run the Davidson algorithm implemented in \textsc{atsp2k} was 11 Gb/proc. Getting the lowest root from the H matrix  took around 22 minutes of CPU time/proc.}. 
To construct the MR-I/CV$\! p$ spaces for $p=20$ and $p=31$, we first omit the CSFs with a \mbox{$\vert c_i \vert < 1.10^{-6}$} in the preceding expansions ($p=14$ and 20), with an impact smaller than 10$^{-6}$ E$_h$ on both energy and $\frac{\hbar^2}{m_e}\Delta$S$_{\text{\sc{sms}}}$. These lists are then completed by adding the CV expansions of configurations $15-20$, and $21-31$, respectively. 

Table \ref{tab:EA} displays smooth convergence trends along its diagonal but the largest expansions used for both S and S$^-$ definitely underestimate the non-relativistic experimental  electron affinity 
($  ^e\!A_{ref}^{NR} = 0.077~026~\mbox{E}_{\mbox{h}} $), indicating that the neutral system is too correlated with respect to the negative ion. We do not report the corresponding table for MR(S$^-$)/MR1(S) models (having 4 columns instead of 15) that displays a good convergence  toward  $  ^e\!A_{ref}^{NR} $.
\begin{table}
\caption{Electron affinity ($^e\!$A, in E$_h$) versus the number of configurations ($p,p'$) in MR2$_p$ for S and MR$_{p'}$ for S$^-$. The absolute energy and total number of CSFs (NCSF) of each model MR-I/CV$\! p$ is given in the first lines and columns of the table. The configurations are taken in the order of increasing weight (see Table \ref{tab:weights}). Underlined are the values of $^e\!$A in reasonable agreement with $  ^e\!A_{ref}^{NR} = 0.077~026~\mbox{E}_{\mbox{h}} $. \label{tab:EA}}
\begin{tabular}{cr|c|cccccc}
\colrule
$p' \backslash p$ && S  & 1 & 2 & 3 &  4  & $\cdots$ & 15 \\
&NCSF && 235 971 & 355 354 & 537 163 & 681 582 &$\cdots$& 2 407 805\\
\colrule
S$^-$	&&	$E$(in E$_h$) & -397.718278 &  -397.722125 &  -397.723707 &  -397.724631 &$\cdots$&  -397.726165 \\
\colrule
1 & 541 780 & -397.794996 & \underline{0.076718} & 0.072871 & 0.071289 & 0.070366 &$\cdots$&  0.068832 \\
3    & 864 954& -397.797229 & 0.078951 & 0.075104 & 0.073522 & 0.072598  &$\cdots$& 0.071064\\
4    & 982 233& -397.797780 & 0.079501 & 0.075654 & 0.074073 & 0.073149  &$\cdots$& 0.071615\\
  5   & 1 088 076 & -397.798944 & 0.080666 & 0.076819 & 0.075237 & 0.074313 &$\cdots$& 0.072779\\
6 &1 210 344&-397.799173 & 0.080895 &\underline{0.077048} &0.075466 & 0.074542 &$\cdots$& 0.073008\\
10 & 2 623 506&-397.799929 & 0.081651 & 0.077804 &0.076222 &0.075298 &$\cdots$& 0.073764\\
12 & 2 854 430&-397.800229 & 0.081951& 0.078104 &0.076522 &0.075598 &$\cdots$& 0.074064\\
14 & 3 175 092&-397.800372 &0.082094 &0.078247 &0.076665 &0.075741 &$\cdots$& 0.074207\\
20\footnote{These lists are truncated (see text). The actual CSF numbers are 2~089~778  and  2~058~776 for $p=20$ and 31, respectively.\label{foot:tabMR2}}
&3 839 474 &-397.800532 & 0.082254 & 0.078407 &0.076825& 0.075901 &$\cdots$& 0.074367\\
31$^\text{\ref{foot:tabMR2}}$ &4 339 910 & -397.800667 &  0.082389 & 0.078542 & \underline{0.076960} & 0.076037 &$\cdots$&  0.074504\\
\colrule
\end{tabular}
\end{table}
For a given correlation model, the $\Delta$S$_{\text{\sc{sms}}}$ parameter is calculated with the wave function expansions that bring the theoretical electron affinity value as close as possible to the $^e\!$A$^{NR}_{ref}$ reference value.  
This approach is supported by the strong correlation observed between the total energy and the S$_{\text{\sc{sms}}}$ parameter, as discussed below. 
When adding configurations, one by one,  in the MR$_p$ of neutral~S, we look for the corresponding model in S$^-$ that gives the best energy balance.  These values are underlined in Table~\ref{tab:EA} and only appear in the first three columns corresponding to MR2-I/CV$\! p$ ($ p = 1-3$), all larger $p \geq 4$ values underestimating the  electron affinity, even for the largest MR-I/CV31 S$^-$ calculation.  The three associated S$^-$  correlation models correspond to MR-I/CV1 (mono-reference), MR-I/CV6 and MR-I/CV31, respectively. 
 For the approach based on the less correlated model for  S (MR1), we select, using the same criteria,  the calculations MR-I/CV4, MR-I/CV6, MR-I/CV12 and MR-I/CV31 of S$^-$ for the four  MR1-I/CV$\! p$ ($ p =1-4$) of S, respectively.

\subsubsection{Valence and Core-valence results}
\label{V_CV_results}

Table~\ref{tab:MRcalc} reports  in two blocks the  electron affinities and S$_{\text{\sc{sms}}}$ differences for the valence correlation models and for their open-core  extensions,  using  the Sulfur MR1 and MR2 models, respectively.
  In the upper half of Table~\ref{tab:MRcalc}, we compare the results of the valence correlation
calculations using the MR1-I$\lceil 9k\rceil$ model (see section~\ref{subsub_MR-MCHF}), with  the values obtained from the four CI calculations based on the core-excited correlation models. In the MR$(p,p')$ adopted notation, $p$ refers to the model used for S while $p'$  refers to S$^-$.
According to our approach, the theoretical  electron affinity values are forced to align with the non-relativistic experimental value  through the $(p,p')$ selection,  but there is no such constraint on  $\Delta$S$_{\text{\sc{sms}}}$. 
Opening the core through the added CV expansion  affects the $\Delta$S$_{\text{\sc{sms}}}$ by up to 12 percents. 
Since the MR1(4,31) calculation corresponds to the complete models, the extracted $\Delta$S$_{\text{\sc{sms}}}$ is {\it a priori} reliable. 
We observed that the results are well aligned, for each system, when plotted in a total energy versus S$_{\text{\sc{sms}}}$ diagram. Furthermore, the relation is similar for  S and S$^-$. A close analysis of the convergence patterns of the MR1($p,p'$) results leads to a 10\% uncertainty estimation on the calculated $\Delta$S$_{\text{\sc{sms}}}$.
 
The second half  of Table~\ref{tab:MRcalc} displays the corresponding results using MR2 for Sulfur. A good consistency with the MR1 $\Delta$S$_{\text{\sc{sms}}}$ values is found for the valence  and first open-core  (MR2(1,1)) models, but  the two larger core-excited CI calculations MR2(2,6) and MR2(3,31) bring unfortunate variations.  The effect of the truncation at $p=3$ of the Sulfur \mbox{MR2-I/CV15} model is estimated by the MR2(4,31) configuration interaction calculation that slightly underestimates the NR  electron affinity. One observes that this extension  affects the $\Delta$S$_{\text{\sc{sms}}}$ value by more than 15\%. Furthermore, the complete calculation MR2(15,31) is unreliable, given its large underestimation of $^e\!$A$^{NR}_{ref}$. 

From all these observations, we  reject the open-core CI models based on MR2. Indeed, the Sulfur model includes much more correlation than the one built for S$^-$. If one goes from a complete model (MR2(15,31)) that strongly underestimates $^e\!$A to a balanced model that  adjusts the  electron affinity, it must be through a too large truncation of the S expansions. The problem of underestimating the S$^-$ correlation energy with respect to the  neutral atom's one is then transferred onto another problem which is the lack of convergence for the latter system (S).

The breakdown of the proposed open-core procedure using the MR2 model for S is probably due to the different nature of the total wave functions obtained for S$^-$ and S. 
The MR1 approach produces  more comparable wave functions for Sulfur and its negative ion, respecting the needed balance. Signs for a large difference between the MR1 and MR2 Sulfur wave function  appear in the analysis of their representation, through the comparison of their respective  spectroscopic orbital mean radii (see Table~\ref{rOrb}) and of their configuration weights (see Table~\ref{tab:weights}). 

\begin{table}[h!]
\caption{Number of CSFs (NCSF), total energy (E, in E$_h$) and S$_{\text{\sc{sms}}}$ parameters (in a$_0^{-2}$) for S and S$^-$.
The last two columns report the corresponding electron affinity ($^e\!$A, in E$_h$) and $\Delta$S$_{\text{\sc{sms}}}$ (in a$_0^{-2}$).
The two sets of results correspond to the zeroth-order multireferences MR1 and MR2 used for S (see text). For each set, the results from the  
valence models (MR-I$\lceil 9k\rceil$) and the open-core CI calculations (MR($p,p')$)  are reported.
\label{tab:MRcalc}}
\begin{tabular}{l|rcc|rcc|cc}
\colrule
& \multicolumn{3}{c|}{S $3p^4 \; ^3P$} & \multicolumn{3}{c|}{S$^-$ $3p^5 \; ^2P^o$} & \multicolumn{2}{c}{$\Delta(\mbox{S}-\mbox{S}^-)$} \\
\multicolumn{1}{c|}{Model} & NCSF & E  & S$_{\text{\sc{sms}}}$  & NCSF & E & S$_{\text{\sc{sms}}}$& $^e\!$A & $\Delta$S$_{\text{\sc{sms}}}$\\
\colrule
MR1-I$\lceil 9k\rceil $ & 43 276 & -397.673394 & -67.01749 & 525 111 & -397.751017 & -67.10600 & 0.07762 & 0.0885\\
MR1(1,4)& 66 280 & -397.720746 & -66.64770 &  982~233 & -397.797780 & -66.73163 & \underline{0.07703} & 0.0839 \\
MR1(2,6) & 181 851  & -397.722224 & -66.61536 & 1 210 344 & -397.799173 & -66.71128 & \underline{0.07695} & 0.0959\\
MR1(3,12) &  268 647 & -397.723184 & -66.60345 & 2 854 430 & -397.800229 & -66.69588 & \underline{0.07705} & 0.0924\\
MR1(4,31) & 408~152 &  -397.723563 & -66.59150 & 4 339 910 & -397.800667 & -66.69335 & \underline{0.07710} & 0.1018\\
\colrule
MR2-I$\lceil 9k\rceil $ & 209 553 & -397.674938 & -67.01897 & 525 111 & -397.751017 & -67.10600 & 0.07608 & 0.0870 \\
MR2(1,1) & 235 971 & -397.718278 & -66.68918 & 541 780 & -397.794996 & -66.78570 & \underline{0.07672} & 0.0965\\
MR2(2,6) & 355 354 & -397.722125 & 	-66.65212 & 1 210 344 &  -397.799173 & -66.71128 & \underline{0.07705} & 0.0592\\
MR2(3,31) & 537 163 & -397.723707 & -66.61143 &4 339 910 & -397.800667 & -66.69335 & \underline{0.07696} & 0.0819\\
MR2(4,31)         & 681 582 & -397.724631 & -66.59715 & &                                & & 0.07604 & 0.0962\\
MR2(15,31) & 2 407 805 & -397.726165 & -66.58303 & & & & 0.07450 & 0.1103\\
\colrule
 NR exp.\footnote{non-relativistic electron affinity $  ^e\!A_{ref}^{NR}$ defined in section~\ref{sub_EA_guideline}.}  & & & & & && 0.07703 &\\
\colrule  
\end{tabular}
\end{table}

\subsection{Theoretical fine structures} \label{fine_structures}

The fine structure splittings are estimated by performing 
Breit-Pauli configuration interaction calculations, including the orbit-orbit interaction. The results are presented in Table~\ref{tab:fine_structures}.
At the Hartree-Fock level of approximation, a large discrepancy between theory and observation ($\simeq$ 30~cm$^{-1}$) is found for
$\; ^2P^o_{1/2-3/2}$ of S$^-$ and $\; ^3P_{1-2}$ of S. Exploring various models for building the zeroth-order non-relativistic wave function, we observe that the inclusion of term-mixing due to $LS$-breakdown does not improve the fine structure splittings. Valence correlation is definitely insufficient to get a satisfactory agreement, as reflected by the splittings reported in~\cite{Froetal:06a}. A ``simple'' correlation model - denoted SD in Table~\ref{tab:fine_structures} -, based on single- and double-excitations up to $6g$  from a single configuration and allowing at most one hole in the $2p$-subshell ({\it i.e.} keeping $1s$ and $2s$ closed), improves significantly the agreement between the theoretical and observed fine structure splitting values. Unfortunately, this agreement is destroyed when progressively extending the reference space to our more elaborate  correlation model, illustrating the difficulty of getting reliable {\it ab initio} fine structure splittings.

\begin{table}[h!]
\caption{Comparison of Breit-Pauli fine structure splittings~(cm$^{-1}$) with observation. \label{tab:fine_structures}}
\begin{tabular}{lccccc}
 \colrule
\multicolumn{1}{c }{ S$^-$}  & \multicolumn{1}{c }{\hspace*{0.5cm} HF \hspace*{0.5cm} } 
        & \multicolumn{1}{c }{\hspace*{0.2cm} SD \hspace*{0.1cm} }
        & \multicolumn{1}{c }{\hspace*{0.2cm} MR-I/CV31 \hspace*{0.1cm} }
        &  \multicolumn{1}{c }{\hspace*{0.1cm} observed~\cite{Bloetal:06a} } \\
  $\; ^2P^o_{1/2-3/2}$ & $-$453.03 & $-$482.07 &  $-$471.16  &  $-$483.54 \\
   \colrule
\multicolumn{1}{c }{S}   & \multicolumn{1}{c }{\hspace*{0.5cm} HF \hspace*{0.5cm} } 
        & \multicolumn{1}{c }{\hspace*{0.2cm} SD \hspace*{0.1cm} }
        & \multicolumn{1}{c }{\hspace*{0.2cm} MR1-I/CV4 \hspace*{0.1cm} }
        &  \multicolumn{1}{c }{\hspace*{0.1cm} observed~\cite{Bloetal:06a} } \\
  $\; ^3P_{1-2}$ & $-$366.91 & $-$394.82   & $-$395.22 & $-$396.06 \\
  $\; ^3P_{0-1}$ & $-$181.89 & $-$174.82 & $-$169.93  &  $-$177.54 \\
 \colrule
\end{tabular}
\end{table}

\section{Isotope shift in the electron affinity: comparison observation - theory}
\label{sec_comp_obs_th}

The observed and theoretical isotope shifts in the Sulfur electron affinity, both determined in the present work, are compared in Table~\ref{tab:results}. 
The observed isotope shift (IS) on the electron affinity of Sulfur is found to be positive, 
$^e\!A$($^{34}$S$)- \; ^e\!A$($^{32}$S) $ = + 0.0023(70)$~cm$^{-1}$, but the uncertainty implies that it may be either ``normal'' 
[$^e\!A$($^{34}$S) $ > \; ^e\!A$($^{32}$S)]
 or ``anomalous'' [~$^e\!A$($^{34}$S) $ < \; ^e\!A$($^{32}$S)].
 The normal mass shift (NMS) is easy to estimate from the first term of eq.~(\ref{NMS_plus_SMS}), 
NMS~$ = 0.016898~\mbox{cm} ^{-1}$, using the observed electron affinity.
 The experimental value of the residual isotope shift (RIS) is obtained by substracting the NMS contribution to the  total isotope shift, {\it i.e.} RIS~$= -$0.0146(70)~cm$^{-1}$.

As far as theory is concerned, two sets of results are reported, omitting or including the core-valence excitations, as described in section~\ref{V_CV_results}. For each set, the electron affinity, the specific mass shift (SMS), the total mass shift (MS = NMS + SMS), the field shift (FS), the total isotope shift (IS = MS + FS) and the residual isotope shift (RIS = IS - NMS) are reported. 
 The electron affinities are compared with the experimental non-relativistic electron affinity ($^e\!A_{ref}^{NR}$) estimated as explained in section~\ref{sub_EA_guideline}, and  reported in the same table. 
The field shift is estimated from equation~(\ref{eq:FS}) by calculating the change in  the electronic densities at the  nucleus $\Delta \rho({\bf 0})_\text{\sc{NR}} $ with the \textsc{density} program~\cite{Boretal:10a}.  
The error bars of the FS values arise from the uncertainty in the root mean squares of the nuclear charge distributions, converted in a 17\% variation of $\delta \langle r^2 \rangle^{AA'}$. 

For the valence calculations, convergence with respect to the number of correlation layers is achieved in both MR1 and MR2 correlation models. The results reported in the ``valence'' line are obtained by averaging the MR1-I$\lceil 9k\rceil $ and MR2-I$\lceil 9k\rceil$ electron affinities and $\Delta$S$_{\textsc{sms}}$ parameters reported in Table~\ref{tab:MRcalc}. Their uncertainty is estimated as half the difference between the two averaged values. The residual isotope shift (RIS = $-$0.0191(3)~cm$^{-1}$) compares satisfactorily with the experimental result \mbox{(IS $-$ NMS) $=-$0.0146(70)~cm$^{-1}$}.

Opening the core is a very difficult task in the MCHF procedure. The presented open-core CI results are limited to the MR1-based models, due to the breakdown of MR2.
The 0.12\% and 10\% error bars reported on the electron affinity and the SMS values, respectively, are estimated from the convergence of the MR1($p,p')$ sequence of results (see Table~\ref{tab:MRcalc}).
The resulting calculated residual isotope shift of $-$0.0222(24)~cm$^{-1}$, remains compatible with the experimental RIS and its error bars, but the theory-observation agreement is tenuous.  
Core excitations affect the SMS contribution by 15\%, increasing the total isotope shift by a large factor (2.4). The field shift constitutes a small fraction ($\sim$ 2\%) of the residual shift (SMS+FS) but  constitutes an important contribution to  the total isotope shift (IS). 

Both theory and experiment agree with a strong cancellation between the specific mass shift (SMS) and the normal mass shift (NMS) contributions.
Although the normal or anomalous character of the isotope shift in the Sulfur electron affinity cannot be strictly confirmed from the present work, the {\it ab initio} calculations are definitely in favour of an anomalous~IS. One should keep in mind however that the theoretical error bars are estimated from an objective analysis of the correlation models but do not take into account core-correlation and relativistic effects that are systematically neglected.
\begin{table}[h!]
\caption{Experimental and theoretical electron affinity, total isotope shift (IS) and residual isotope shifts (RIS). For the {\it ab initio} calculations, the specific mass shift (SMS), the total mass shift (MS) and the field shift (FS) contributions are reported separately. All values in $10^{-2} \mbox{cm}^{-1}$.
  \label{tab:results}}
\begin{tabular}{llcccccc}
 \colrule
   & \multicolumn{1}{c }{ $^e\!$A \hspace*{0.5cm} } 
        & 
        & 
        & 
        &  \multicolumn{1}{c }{\hspace*{0.1cm} IS } 
        &  \multicolumn{1}{c }{\hspace*{0.1cm} RIS } \\
 \colrule
 \multicolumn{7}{c}{observation} \\
  exp. & 1675298 &  & & & $+$0.23(70) & $-$1.46(70) \\
 NR exp.\footnote{non-relativistic electron affinity $  ^e\!A_{ref}^{NR}$ defined in section~\ref{sub_EA_guideline}.} & 1690524 &&&&\\
 \multicolumn{7}{c}{theory\footnote{adopting the experimental NMS.} } \\
   &
        & \multicolumn{1}{c }{\hspace*{0.7cm} SMS \hspace*{0.7cm} }
        & \multicolumn{1}{c }{\hspace*{0.7cm} MS  \hspace*{0.7cm} }  
        & \multicolumn{1}{c }{\hspace*{0.7cm} FS \hspace*{0.7cm} }  
& & \\
 \colrule
valence\footnote{averaging the MR1-I$\lceil 9k\rceil $ and MR2-I$\lceil 9k\rceil $ results~(see text).} & 16867(169)~10$^{+2}$ & $-$1.94(2)& $-$0.25(2) &0.036(6)
& $-$0.22(3) & $-$1.91(3) \\
+ core-valence\footnote{using the MR1(4,31) results of Table~\ref{tab:MRcalc}} & 16922(200)~10$^{+2}$ & $-$2.25(23) & $-$0.56(23) & 0.038(7) & $-$0.53(24) & $-$2.22(24) \\
 \colrule
\end{tabular}
\end{table}

\section*{Acknowledgements}

 T.~Carette and M.~Godefroid thank the Communaut\'e fran\c{c}aise of Belgium (Action de Recherche Concert\'ee)
and the Belgian National Fund for Scientific Research (FRFC/IISN Convention) for financial support. They are grateful to Georges Destr\'{e}e of the ULB-VUB Computing Centre for his help in adapting {\textsc{atsp2k} on the {\sc Hydra} cluster.
\noindent
%

\newpage

 \begin{figure}
	\begin{center}
		\includegraphics[scale=0.5,angle=0]{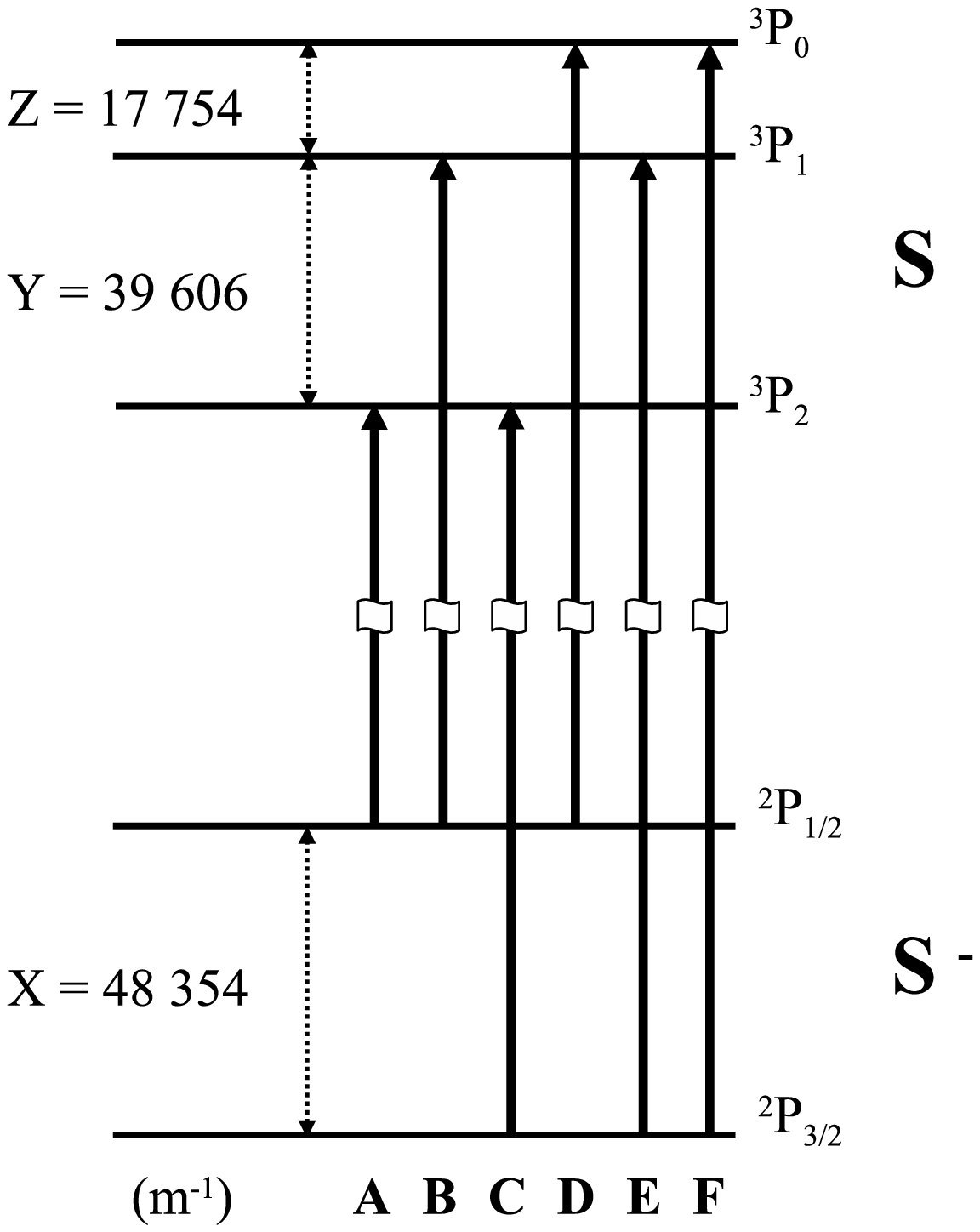}
		\caption{S$^{-}$ fine structure detachment thresholds. Energies measured with wavenumbers in m$^{-1}$. (figure taken from \cite{Bloetal:06a}).}
	\label{fig:FS_thresholds}
	\end{center}
\end{figure}

\clearpage

\begin{figure}
	\begin{center}
		\includegraphics[scale=0.5,angle=0]{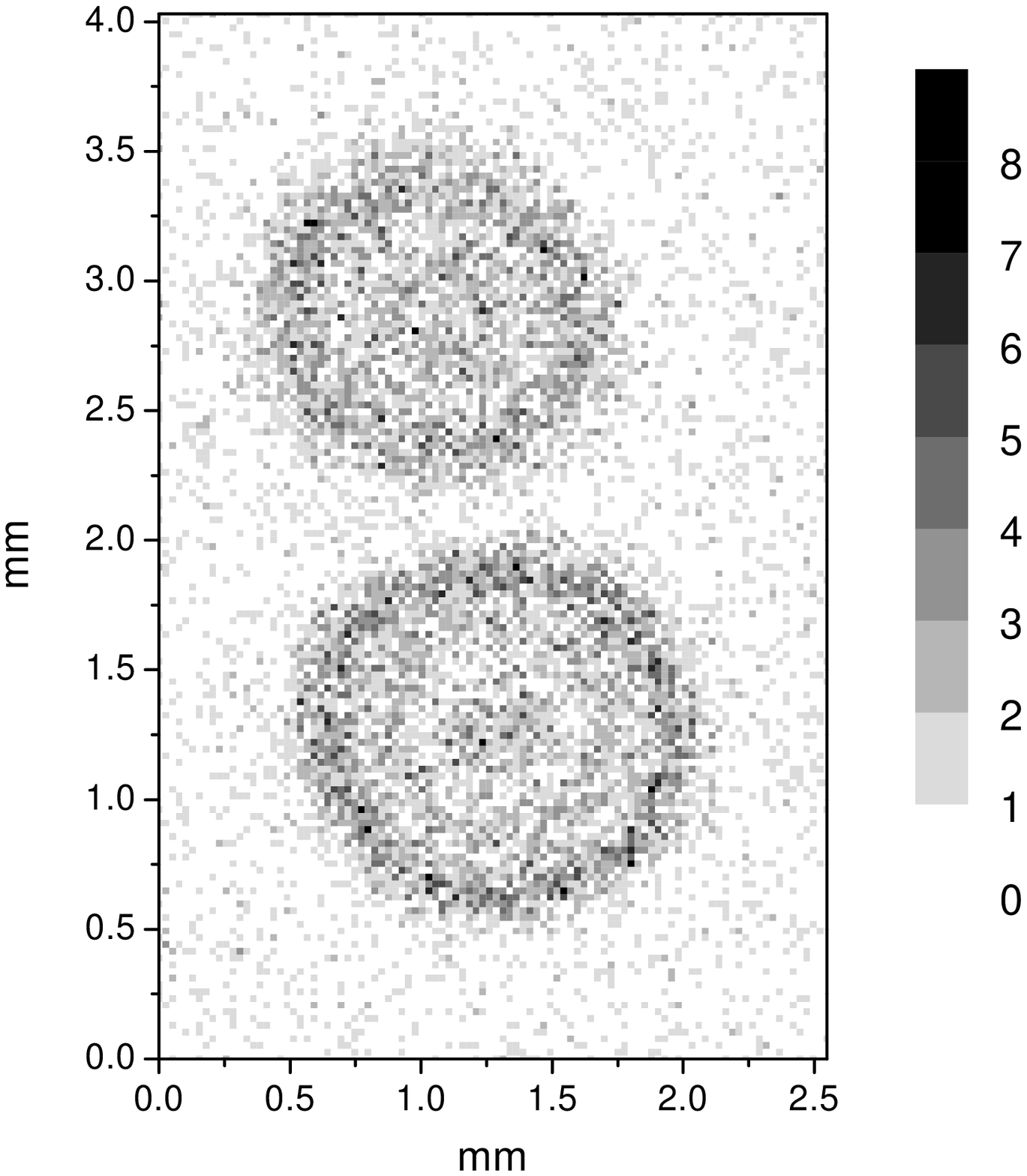}
		\caption{Double interferogram obtained from $^{34}$S$^-$ at a wavelength $\lambda$=596.89056(2)~nm, in an electric field 291~V/m, for an accumulation time of 2000~s. The grey scale indicates the total number of electrons counted in each pixel. The data are recorded by means of a Quantar Technology Inc. particle detector of the series 3391 with the 2251 Image-Trak\texttrademark enhanced software. The presented image was reprocessed with Microcal\texttrademark Origin\textregistered.}
	\label{fig:interferogram}
	\end{center}
\end{figure}

\begin{figure}
	\begin{center}
		\includegraphics[scale=0.5,angle=-90]{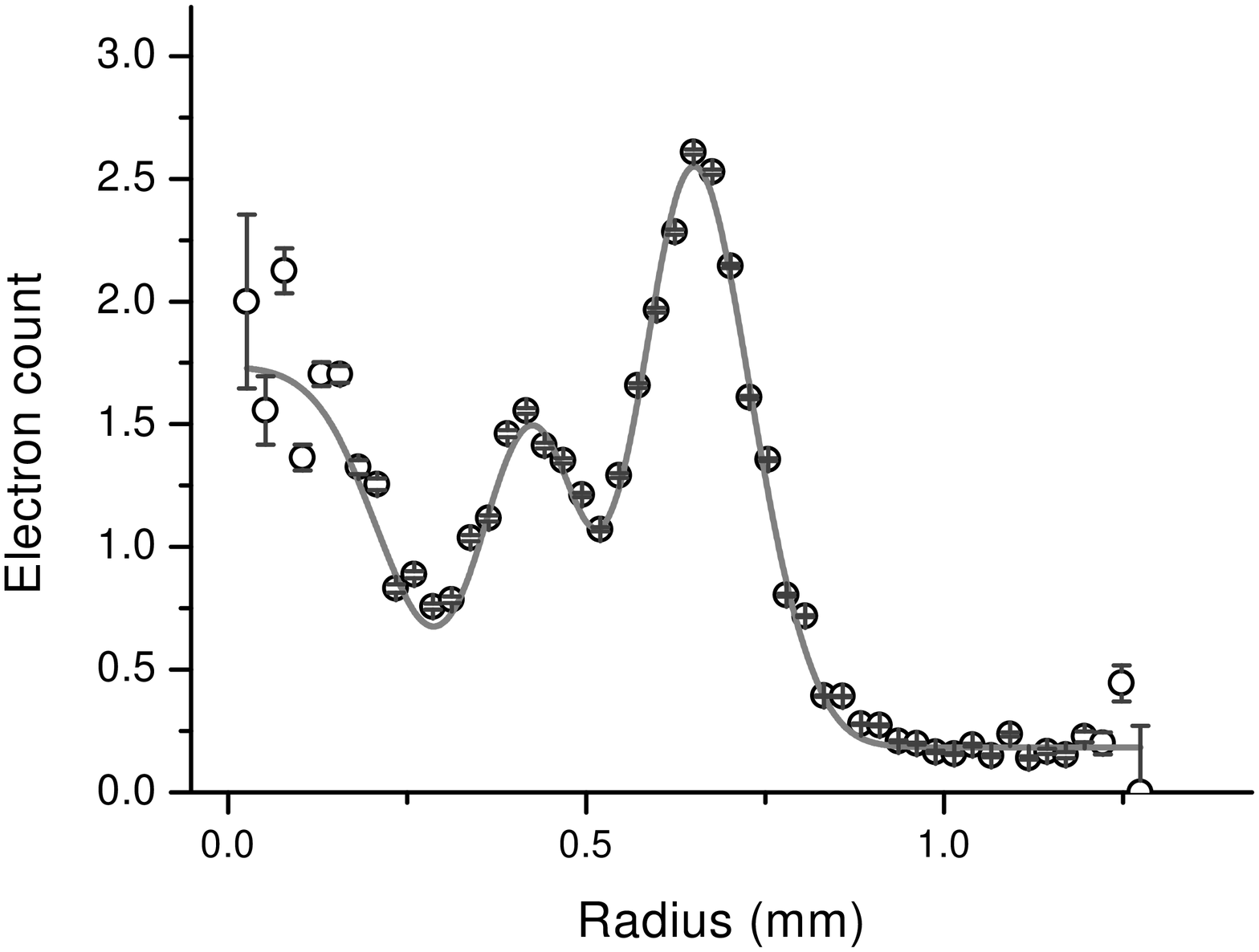}
		\caption{Average number of electrons counted per pixel, in the lower spot of Fig.~\ref{fig:interferogram}. The continuous line is the radial profile calculated with the best-fitting parameters (as provided by an adjustment algorithm applied to the original 2D data). One among these parameters is the initial kinetic energy of the electron, here found to be 0.5819(37) cm$^{-1}$ (but the $1\sigma$ error bar given here assumes no uncertainty at all on the electric field itself).}
	\label{fig:Radprof}
	\end{center}
\end{figure}

\begin{figure}
	\begin{center}
		\includegraphics[scale=0.45,angle=-90]{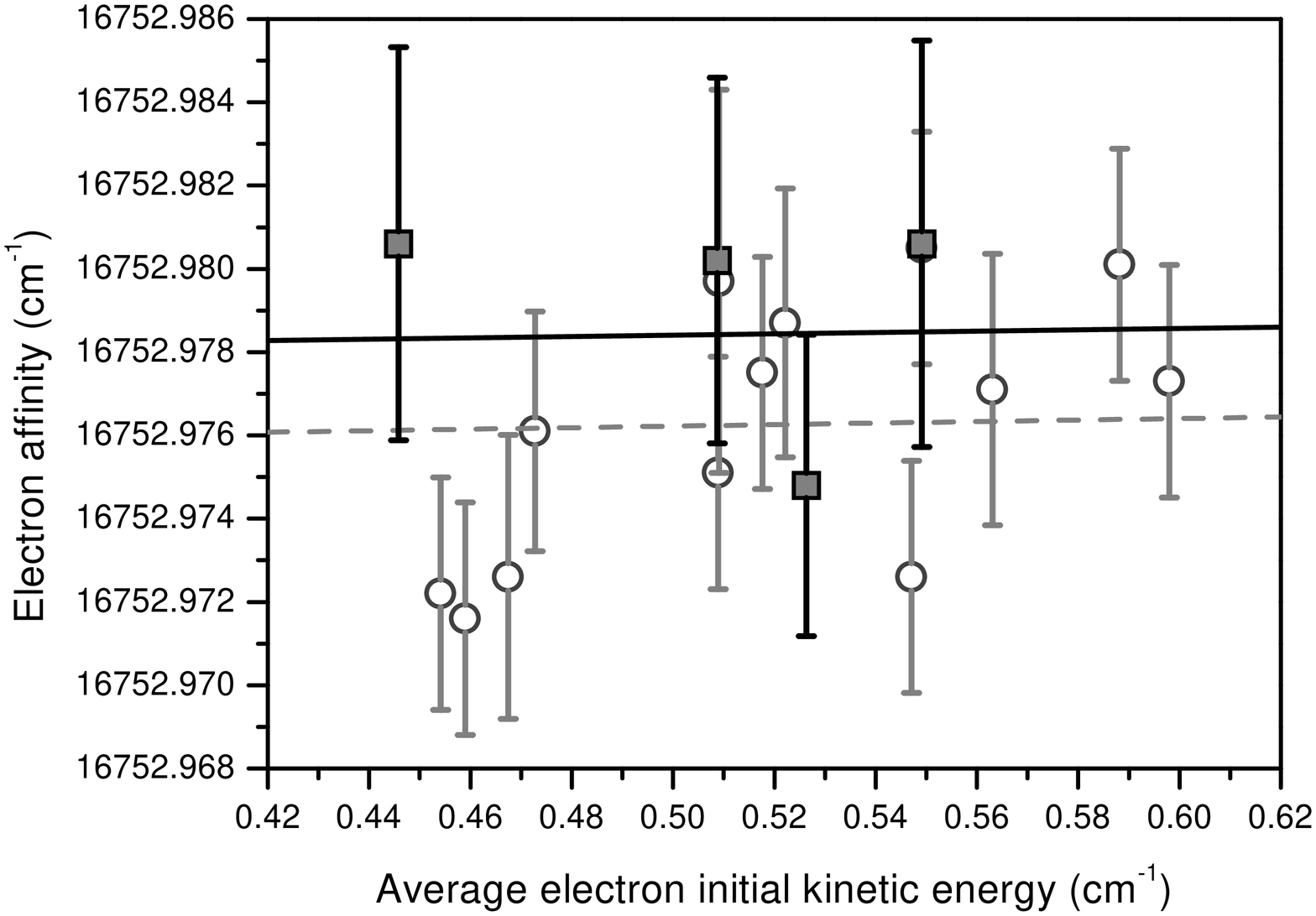}
		\caption{Comparison of electron affinity measurements made in common series of experiments for $^{32}$S (circles) and $^{34}$S (squares), with the average trend for each isotope (dashed line is 32, continuous line is 34), assuming similar dependences of the apparent electron affinity as a function of the average electron kinetic energy. The data shown in Fig.~\ref{fig:interferogram} produce the experimental $^{34}$S point at 0.509 cm$^{-1}$.}
	\label{fig:donnees}
	\end{center}
\end{figure}
\end{document}